\documentclass[twocolumn,preprintnumbers,amsmath,amssymb]{revtex4}
\usepackage{mathrsfs}
\usepackage{amssymb}
\usepackage{graphicx}
\usepackage{dcolumn}
\usepackage{bm}
\usepackage{textcomp}
\usepackage{amsmath}
\usepackage[titletoc]{appendix}

\newcommand\ket[1]{\ensuremath{|#1\rangle}}

\newcounter{RomanNumber}

\begin{document}
\title{Effective Eavesdropping to Twin-Field  Quantum Key Distribution
}
\author{Xiang-Bin Wang,$ ^{1,2,4\footnote{Email
Address: xbwang@mail.tsinghua.edu.cn}\footnote{Also a member of Center for Atomic and Molecular Nanosciences at Tsinghua University}}$ Xiao-Long Hu$ ^{1}$,and  Zong-Wen Yu$ ^{3}$
}

\affiliation{ \centerline{$^{1}$State Key Laboratory of Low
Dimensional Quantum Physics, Department of Physics,} \centerline{Tsinghua University, Beijing 100084,
People¡¯s Republic of China}
\centerline{$^{2}$ Synergetic Innovation Center of Quantum Information and Quantum Physics, University of Science and Technology of China}
\centerline{  Hefei, Anhui 230026, China
 }
\centerline{$^{3}$Data Communication Science and Technology Research Institute, Beijing 100191, China}
\centerline{$^{4}$ Jinan Institute of Quantum technology, SAICT, Jinan 250101,
People¡¯s Republic of China}}
\begin{abstract}
We present an effective Eavesdropping scheme to attack the twin-field protocol of quantum key distribution (TF-QKD)
proposed recently (M. Lucamarini, Z. L. Yuan, J. F. Dynes£¬ and A. J. Shields, Nature, 2018). After carrying out our scheme, Eve will have full information of all bit values due to signal fields without causing any noise. In such a case, the actual secure key rate is 0 but the existing protocol of TF-QKD will present a key rate of ${50\%}$ from raw key to final key. This shows that the existing protocol of TF-QKD is supposed to be further studied  so as to see whether its security can be improved to a satisfactory level. In view of this security problem of the TF-QKD, the 4-intensity optimized protocol  of measurement-device-independent QKD (Phy. Rev. A {\bf{93}}, 042324 (2016) ) is still the one that can present the longest secure distance of QKD on the Earth.

\end{abstract}


\maketitle


\section{Introduction}\label{Sec:Intro}
Quantum key distribution (QKD) [1,2] can be used for secure private communication in principle. But in practice, there can be security loopholes due to device imperfection. Since the concept of measurement device independent quantum key distribution (MDI-QKD)[3,4] was proposed, the fatal security threaten of detection attack [5] was thoroughly resolved. Actually, the MDI-QKD is secure even though Eve. completely controls the channel and the measurement station. Later, many efforts were made to improve the practical feasibility of MDI-QKD[6-17]. In particular, Ref.[6] for the first time present the operable method for the decoy-state MDI-QKD with finite intensities so that the protocol is experimentally operable in practice[7]. The theoretical method in Ref.[13] can hugely improve the key rate and secure distance. There, it proposes to use the 4-intensity optimization protocol with joint constraints and economic worst-case analysis. This method [13] has been experimentally verified extensively[14-17]. Specifically, using this method, a secure distance of 404km was experimentally demonstrated[15]. This is the longest secure distance of QKD on the earth.

Very recently,  the twin-field quantum key distribution (TF-QKD)[18] was proposed with a claim of over 500km accessible distance. However, as was stated in [18], the security proof is not completed because in the protocol Alice and Bob will have to announce the phase information of their pulses, this does not meet the condition of the decoy-state method[6,19,20,21]. Here we show that Eve. can indeed attack the TF-QKD effectively by making use of the phase information announced by Alice and Bob after the measurement and classical communication as requested by the protocol. In what follows, we shall first review the main ideas of the TF-QKD and Eavesdropping in Section II, and then present our Eavesdropping scheme in Section III. After that we show details of the consequence of our Eavesdropping scheme in Section IV and finally we conclude our paper in Section V.
\section{Main idea of TF-QKD and Eavesdropping}
Similar to the measurement-device-independent quantum key distribution(MDI-QKD)[3,4], in the twin-field quantum key distribution (TF-QKD)[18], Alice and Bob sends field to the un-trusted third party Charlie. MDI-QKD is secure even Eve controls the whole channel and Charlie.  In and virtual  protocol, Alice and Bob initially shares single-photon entangled states of $|\psi^+\rangle=\frac{1}{\sqrt 2}(|01\rangle+|10\rangle)$. Alice and Bob will do a phase shift of either 0 or $\pi$ and they will send their fields to Charlie. After a collective measurement, Charlie will see whether the single-photon bipartite state is $|\psi^+\rangle$ or $|\psi^-\rangle=\frac{1}{\sqrt 2}(|01\rangle-|10\rangle)$. Although this information is announced publicly, no one knows which value, 0 or $\pi$ was selected by Alice or Bob in doing their phase shift, although it is known to everyone whether Alice and Bob has used the same phase shift or different phase shift.

However, in a real TF-QKD protocol, we do not have such an initially shared states. The TF-QKD proposed to use weak coherent states at each side. In what follows we show that the TF-QKD is actually unsecure if Eve controls the whole channel and Charlie. As it is stated in Ref[18], the security
 proof of TF-QKD is not completed because  it has not considered the  influence to security of the final announcement of the random phase values taken by Alice and Bob. In our proposed Eavesdropping scheme, we make use of this and Eve. does operations after the announcement. Also, before that, in doing the interference measurement, Eve. uses  crude measurement which only collapses the light beams into two subspaces, vacuum or non-vacuum. If a beam is collapsed to non-vacuum subspaces, it is a linear superposition of different Fock state and it keeps the phase difference of different photon-number states. Also, Eve. will engineer this non-vacuum linear superposed state by non-trace-preserving maps. If she obtains the outcome she wants, she will announce to Alice and Bob she has obtained a count and stores the outcome light state. If she has not obtained the outcome she wants, she  just pretends that she has not detected anything. After the random phase shift values taken earlier by Alice and Bob are announced, Eve does projective measurement to her stored light state corresponding to the signal fields again and the outcome  presents deterministically all bit information.

Recall the TF-QKD protocol in Fig.[1]. The protocol post selects those events whose random phases taken by Alice and Bob are in the same slice. In the limit of very small slice, only those pulse pairs with same random phases will be post selected. For simplicity, we shall only consider such a case. Or equivalently, we shall consider an ideal protocol where Alice and Bob always take the identical random phase shifts, corresponding to infinitely small phase slices in [18].   We shall only consider such an "ideal TF-QKD protocol". We assume that there is no operational or detection errors.
\begin{figure}
    \includegraphics[width=200pt]{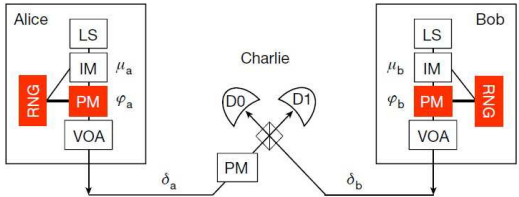}
    \caption{Schematic protocol of TF-QKD  proposed by [18]. This picture is taken from Ref.[18]}
\end{figure}

\section{Eavesdropping scheme}
 We list the Eavesdropping scheme to each pulse pairs from Alice and Bob in the following:
\\{\bf Step 0}.
 Consider Fig.1. Before the twin pulses enter the beam splitter, Eve.(Charlie) just honestly does whatever as requested by the the TF-QKD protocol.
 \\ {\bf Step 1}. Eve. only takes crude detection at the measurement phase after a pulse pair passes the beam splitter in Fig.1. Say, if detector $D0$ or $D1$ counts, it means the the corresponding output beam has been collapsed to the Fock space without vacuum. But Eve. does not immediately announce her detected result as requested by the protocol at this moment. She put off her announcement until the end of {\bf Step 4} in the attacking scheme. {\em This is the key point of the attacking scheme.}
 If she obtains vacuum, she stops the attacking and she will announce that she has detected nothing for the corresponding incident pulse pair from Alice and Bob at the end of {\bf Step 4}; and if she obtains non-vacuum, she stores the detected state $|\chi\rangle$ and continue the attacking scheme.

 Remark: Consider the beam at any output port of the beamsplitter. Given a linear superposed state in the form $\sum_{k=0}^\infty a_k|k\rangle$ just before it comes into the corresponding detector. If the detector counts then, the outcome state becomes into \begin{equation}
 |\chi\rangle=\mathcal{N}\sum_{k=1}^\infty a_k|k\rangle,
 \end{equation} where $\mathcal N$ is the normalization factor.
 \\{\bf Step 2}. Eve takes a crude measurement to project the stored state $|\chi\rangle$ either to the subspace $\mathcal S=\{|1\rangle, |2\rangle\}$ or to the subspace $\tilde{\mathcal S}=\{|3\rangle,|4\rangle,|5\rangle\cdots \}$. If the outcome is $\tilde{\mathcal S}$, she stops attacking and she will announce that she has detected nothing for the corresponding incident pulse pair from Alice and Bob at the end of {\bf Step 4}; and if the outcome is $\mathcal S$, she stores the state and continues her attacking scheme.

 Remark: If she obtains subspace $\mathcal S$, the stored state has become to
 \begin{equation}
 |\chi_2\rangle = \frac{1}{\sqrt {|a_1|^2+| a_2|^2}}(a_1|1\rangle+a_2|2\rangle).
 \end{equation}
\\Step 3. Eve. takes the following unitary transformation to her stored state:
\begin{align}\label{equ:U1}
    U_{1}:\left\{ \begin{array}{l}
    \ket1 \rightarrow \sqrt {\mu} \ket1 +\sqrt{1-\mu}\ket{m_0} \\
    \ket2 \rightarrow \ket2
    \end{array} \right.
\end{align}
where $\mu$ is the intensity of signal state pulse in the TF-QKD protocol and   $\ket{m_0}$ is a state orthogonal to both $\ket1$ and $\ket2$.

Remark: Same with all existing decoy-state protocol, the intensities of all signal pulses and decoy pulses are determined prior to the protocol and they are all known to Eve.
\\{\bf Step 4}. Eve. takes crude measurement to the stored state after {\bf Step 3} and the state will be collapsed to either state $|m_0\rangle $ or the subspace $\mathcal S$ spanned by the Fock states $\{|1\rangle,|2\rangle\}$ as introduced in {\bf Step 2}. If she obtains $|m_0\rangle$, she stops attacking and she  announces that she has detected nothing; and if she obtains subspace $\mathcal S$, she stores the state and announces which detector ($D0$ or $D1$) has counted.

Remark: After {\bf step 4}, the stored state is
\begin{equation}
|\chi_4\rangle = \mathcal N_4 (a_1\sqrt {\mu} |1\rangle + a_2|2\rangle ).
\end{equation}
\\Step 5. After Alice and Bob announce the information of their random phase shift, bases of each pulse pairs, and which pulses are decoy pulses and which pulses are signal pulses,  Eve. takes a phase shift operation and her stored state is changed into the following state if the corresponding incident pulse pair is a signal pair
\begin{equation}
|\chi_5\rangle = \frac{1}{\sqrt 2}(|1\rangle\pm |2\rangle).
\end{equation}
depending on the the relative phase between $a_1$ and $a_2$. This enables Eve. to know the bit value for sure without causing any noise.
does a projective measurement to those stored states corresponding to signal pulse pairs and she will be sure the corresponding bit values. We can see details in the following section.

\section{ State evolution along with Eavesdropping}  Charlie's beam splitter   performs a unitary transformation in the form(Fig.1):
\begin{align}\label{equ:BS}
    U_{BS}:\left\{ \begin{array}{l}
    a^\dag \rightarrow (a^{\dag} + b^{\dag})/\sqrt{2}\\
    b^\dag \rightarrow (a^{\dag} - b^{\dag})/\sqrt{2}
    \end{array} \right.
\end{align}

In our paper we shall always define modes $a$ or $b$ by propagation direction. In the figure we have assumed the incident mode from Alice's side to be mode $a$, and the one from Bob's side to be mode $b$.

Define the coherent state $|\alpha\rangle = e^{-|\alpha|^2/2} \sum_{k=0}^\infty \frac{\alpha^k}{\sqrt k!}  |k\rangle$.
Given the beamsplitter defined above, any incident pulse pair in  the two-mode coherent state of the form
$|\alpha\rangle_a|-\alpha\rangle_b$, detector $D0$ is the only one that can count. Also, if the incident pulse pair is in the two-mode coherent state of the form
$|\alpha\rangle_a|\alpha\rangle_b$, detector $D1$ is the only one that can count. Here the different subscripts ($a,b$) indicate different modes. (And later for simplicity we shall omit the subscripts and we always use the first ket for mode $a$ and the second ket for mode $b$. )
Eve.'s task is,  in the case that detector $D0$ or $D1$ counts, if the incident pulse pair in {\bf Step 1} is a signal pair, what is the corresponding bit value ?
We consider the case that detector $D0$ counts at Step 1 of Eavesdropping scheme. If the corresponding incident pulse pair
before passing through the beamsplitter was a signal pair (in Z basis), we have $|\alpha|=\sqrt\mu$ and the corresponding state of the beam pair incident to Charlie's beamsplitter in {\bf Step 1} can be either
\begin{equation}
|\psi\rangle = |\sqrt \mu e^{i\rho}\rangle|-\sqrt\mu e^{i\rho}\rangle
\end{equation}
for bit value 0 or
\begin{equation}
|\psi'\rangle = |-\sqrt \mu e^{i\rho}\rangle|\sqrt\mu e^{i\rho}\rangle
\end{equation}
for bit value 1. Here $\rho$ is the random phase shift taken earlier by Alice and Bob. Note that the strong reference light is with Eve., we have just set the phase of reference light to be 0 for simplicity. We can easily calculate the time evolution of each states. For the evolution of state $|\psi\rangle$ (or $|\psi'\rangle)$, we use notation $|\psi_i\rangle $ (or $|\psi_i'\rangle$)  for the non-vacuum state stored by Eve. at the end of Step i of Eavesdropping scheme. Also, since the output mode $a$ (the output mode at the side of  detector $D1$ ) is always vacuum, we only consider time evolution of the output port  mode $b$ (the one at the side of detector $D0$).  Given the incident state $|\psi\rangle$, we have the following states for its time evolution at each stages:
\begin{equation}
|\psi_1\rangle = \mathcal N_1 (|\sqrt {2 \mu} e^{i\rho}\rangle - e^{-2\mu}|0\rangle
=\mathcal N_1 \sum_{k=1}^\infty \frac{(\sqrt {2\mu}e^{i\rho})^k}{\sqrt{k!}}|k\rangle;
\end{equation}
\begin{equation}
|\psi_2\rangle = \mathcal N_2 (\sqrt{2\mu}e^{i\rho}|1\rangle+\sqrt 2\mu e^{2i\rho}|2\rangle);
\end{equation}
\begin{equation}
|\psi_4\rangle = \frac{1}{\sqrt 2} (|1\rangle+e^{i\rho}|2\rangle).
\end{equation}
All parameters $\mathcal N_1,\mathcal N_2,\mathcal N_4$ are normalization factors. After Eve. knows the value of random phase $\rho$, she can take a phase shift of $e^{-i\rho}$ to every photon and obtain:
 \begin{equation}
 |\psi_5\rangle = \frac{1}{\sqrt 2}(|1\rangle + |2\rangle).
 \end{equation}
 Similarly, given the incident state $|\psi'\rangle$ in {\bf Step 1}, we have the following states for its time evolution at each stages:
\begin{equation}
|\psi_1'\rangle = \mathcal N_1 \sum_{k=1}^\infty \frac{(-\sqrt {2\mu}e^{i\rho})^k}{\sqrt{k!}}|k\rangle;
\end{equation}
\begin{equation}
|\psi_2'\rangle = \mathcal N_2 (-\sqrt{2\mu}e^{i\rho}|1\rangle+ \sqrt 2\mu e^{2i\rho}|2\rangle);
\end{equation}
\begin{equation}
|\psi_4'\rangle = \frac{1}{\sqrt 2} (-|1\rangle+e^{i\rho}|2\rangle)
\end{equation}
and
\begin{equation}
 |\psi_5'\rangle = \frac{1}{\sqrt 2}(-|1\rangle + |2\rangle).
 \end{equation}
 A projective measurement can distinguish state $|\psi_5\rangle$ and state $|\psi_5'\rangle$ deterministically.
 In the same way, one can easily show that Eve. can also obtain full information of bit values for those signal pairs if
 detector $D1$ counts.
 \section{Concluding remark.}
 After the Eavesdropping above, Eve. has full information of all raw bits. While to Alice and Bob, if they go ahead to distill the final key using the decoy-state method, they will find a key rate of $50\%$ from raw key to final key. In the Eavesdropping above, the fraction of bits caused single-photon state is $50\%$ among all raw bits. And in the scheme we have assumed the ideal condition there is no error. According to key rate formula (Eq.(2)) of Ref.[18], TF-QKD will present a key rate of $50\%$ from raw key to final key. This means that the existing form of TF-QKD protocol is not secure under our designed Eavesdropping scheme we have presented and the security of the existing  TF-QKD is actually not equivalent to that of the decoy-state MDI-QKD. Therefore, the TF-QKD  is supposed to be further studied in the future to see whether its security can be improved to a satisfactory level. For example, under the attacking scheme we have proposed here, Alice and Bob use many or two significantly different intensities for signal pulse pairs could help to raise the security level. But in such a case the security problem is not trivial. Eve can have many other choices, e.g., she can intentionally announce some wrong results. The security problem for the TF-QKD  also concludes that  the 4-intensity optimized protocol [13] of MDI-QKD is still the one that can present the longest secure distance of QKD without using quantum memory on the Earth [15]. Obviously, using the idea of our Eavesdropping scheme, there are lots of refined variants. This will be reported later.

\end{document}